# Spin-Valley Locking Effect in Defect States of Monolayer MoS$_2$

Yaqian Wang[1,2], Longjiang Deng[1,2], Qilin Wei[3], Yi Wan[4], Zhen Liu[1,2], Xiao Lu[1,2],Yue Li[1,2], Lei Bi[1,2], Li Zhang[1,2], Haipeng Lu[1,2], Haiyan Chen[1,2], Peiheng Zhou[1,2], Linbo Zhang[1,2], Yingchun Cheng[3,*], Xiaoxu Zhao[5,*], Yu Ye[4], Wei Huang[3], Stephen John Pennycook[5], Kian Ping Loh[6] and Bo Peng[1,2,*]

[1] National Engineering Research Center of Electromagnetic Radiation Control Materials, School of Electronic Science and Engineering, University of Electronic Science and Technology of China, Chengdu, 611731, China

[2] State Key Laboratory of Electronic Thin Films and Integrated Devices, University of Electronic Science and Technology of China, Chengdu, 611731, China

[3] Key Laboratory of Flexible Electronics & Institute of Advanced Materials, Nanjing Tech University, Nanjing, 211816, China

[4] State Key Laboratory for Artificial Microstructure & Mesoscopic Physics, Frontiers Center for Nano-optoelectronics, School of Physics, Peking University, Beijing 100871, China

[5] Department of Materials Science and Engineering, National University of Singapore, 9 Engineering Drive 1, 117575, Singapore

[6] Department of Chemistry and Centre for Advanced 2D Materials and Graphene Research Centre, National University of Singapore, 117549, Singapore

* To whom correspondence should be addressed. Email address: bo_peng@uestc.edu.cn (B.P.); iamyccheng@njtech.edu.cn (Y.C.C); xiaoxu_zhao@u.nus.edu (X.X.Z)

**Abstract:** Valley pseudospin in two-dimensional (2D) transition-metal dichalcogenides (TMDs) allows optical control of spin-valley polarization and intervalley quantum coherence. Defect states in TMDs give rise to new exciton features and theoretically exhibit spin-valley polarization; however, experimental achievement of this phenomenon remains challenges. Here, we report unambiguous valley pseudospin of defect-bound localized excitons in CVD-grown monolayer $MoS_2$; enhanced valley Zeeman splitting with an effective *g*-factor of -6.2 is observed. Our results reveal that all five *d*-orbitals and the increased effective electron mass contribute to the band shift of defect states, demonstrating a new physics of the magnetic responses of defect-bound localized excitons, strikingly different from that of A excitons. Our work paves the way for the manipulation of the spin-valley degrees of freedom through defects toward valleytronic devices.

**Keywords:** valleytronic, spintronic, spin manipulation, defect engineering, defect exciton

Manipulation of the spin degree of freedom (DOF) at the atomic scale is crucial in exploring new spintronic and valleytronic devices for quantum information and communication technologies.[1-6] Defects originating from zigzag edge, dislocations and deformations have been experimentally and theoretically demonstrated to lead to long-range magnetic order in transition-metal dichalcogenides (TMDs);[7-10] the proton irradiation induce vacancy defects in $MoS_2$ bulk, giving rise to ferromagnetism.[11] Theoretical studies have also predicted that magnetic atom substitutional doping can introduce magnetic ordering in TMDs;[12-15] and recent experimental works have demonstrated that Zeeman splitting of valley pseudospin is enhanced by Co and Fe atom doping in $MoS_2$ monolayer.[16-17] These create opportunities for spin control and manipulation in TMDs by defects. In parenthesis, defects in 2D materials can more strongly trap free carriers and localize excitons and influence their physical properties, opening up plenty of opportunities to tailor the transport and optical properties of 2D materials. Dopant defects modify the electronic structures of monolayer TMDs, resulting in a shift in the PL peak energy;[18] vacancy defects induce a new low-energy suboptical gap inside the bandgap and lead to a sub-bandgap emission from defect-bound localized excitons.[19-20] Single photon emission in two-dimensional materials due to single defect[21-25] and nanoscale strain,[26-27] particularly in monolayer $WSe_2$ and BN,[28-33] have also been demonstrated.

Coupling of the broken spatial inversion symmetry and strong spin-orbit coupling (SOC) leads to valley pseudospin in monolayer TMDs, in which spin-up holes and spin-down electrons reside in the +K and –K valleys.[34] The separated valleys in momentum space can be regarded as a unique DOF for information coding and transmission. Most recently, spin-valley polarization of defect states in monolayer TMDs has been theoretically predicted.[19, 35] The presence of defects in monolayer TMDs leads to quenching of the circularly polarized photoluminescence (PL) emission of A and B excitons and degradation of their valley polarization due to the increased nonradiative decay rate.[36] Magnetic field can break the valley degeneracy and tune the valley polarization of defect states.[37] Although the emission of defect-bound excitons attributed to chalcogen vacancies has been studied in monolayer $MoS_2$, $MoSe_2$, $WS_2$ and $WSe_2$.[20, 38-41] However, to data, experimentally manipulating the valley pseudospin and splitting of defect states still remains challenging.

In this work, we report observation of valley pseudospin of defect-bound excitons in CVD-grown monolayer $MoS_2$ by circularly polarized PL spectroscopy; a strong polarization-resolved defect emission is observed at 10 K. Remarkably, the valley pseudospin of defect states can be lifted by a magnetic field, analogous to valley Zeeman splitting of intrinsic A excitons of monolayer $MoS_2$, giving an effective *g*-factor of approximately -6.2, which is consistent with the theoretical *g*-factor of -6.6. We proposed that the valley pseudospin and valley Zeeman splitting of defect excitons originate from the momentum-dependent carrier distribution of defect states in the vicinity of ±K valleys; the enhanced *g*-factor arises from the increased effective electron mass and *d*-orbital magnetic moment.

Monolayer $MoS_2$ belongs to the nonsymmorphic space group ($D_{6h}^4$ symmetry and $P6_3/mmc$ space group). In each unit cell, one Mo atom is coordinated by six S atoms and sandwiched between S atoms, forming a S-Mo-S trigonal prismatic geometry, which further shares vertexes to build a hexagonal closed-packed network (Fig. 1a and b). We studied the electronic band structures of monolayer $MoS_2$ in the presence of with one single S vacancy ($V_S$) and S divacancy ($V_{2S}$) by first-principles calculations.[42-43] Pristine monolayer $MoS_2$ without any defect exhibits a distinguishable carrier distribution in two momentum-dependent valleys; spin-up holes and spin-down electrons are only kept in the valance and conduction band edge in the +K and –K valleys (Supplementary Information Fig. S1c).[2-3, 34] Importantly, the defect states also have a unique electronic band structure with two momentum-dependent band minima at the +K and –K points of the Brillouin zone (Fig. 1c). The defect band edges of both the $V_S$ and $V_{2S}$ have spin-up and spin-down states in the +K and –K valleys, respectively, while the valence band edges show opposite spin states in the same valley with defect states, suggesting valley pseudospin of defect states and distinct momentum-dependent carrier distribution (Fig. 1d and Supplementary Information Fig. S1).

Structural defects can be produced in TMDs through the CVD method, such as grain boundaries, various vacancy defects and substitutional defects.[44-45] Moreover, chalcogenide vacancies easily form due to their low formation energy and the high volatility of chalcogenides.[15] Therefore, we designed experimental conditions to grow defect-engineered monolayer $MoS_2$ on Si/SiO$_2$ substrates by the CVD method (See Methods).[46] The Raman frequency difference between the $E_{2g}$ and $A_{1g}$ modes is ~20

cm$^{-1}$ (See Supplementary Information Fig. S2), which satisfies the criteria of CVD-grown monolayer MoS$_2$ on SiO$_2$/Si substrates.[47] To validate the S vacancies, we investigated the structure of the CVD-grown monolayer MoS$_2$ on an atomic scale by employing an aberration-corrected scanning transmission electron microscope with an annular dark field detector (STEM-ADF) operated at low accelerating voltage (60 kV). The maximum transferred energy from 60 kV electron beam is lower than the threshold of the knock on damage of MoS$_2$, thus the defects cannot be induced by electron beam. From the top view, monolayer MoS$_2$ is a hexagonal closed-packed structure; one Mo atom is surrounded by three S atoms, and *vice versa*. Only V$_S$ and V$_{2S}$ defects are observed. The location and density of V$_S$ and V$_{2S}$ is reflected by the contrast reduction in the STEM-ADF image. The mapping of all atom species of the MoS$_2$ monolayer is depicted in Fig. 2a. Selected V$_{2S}$ are highlighted by white dashed circle. Numerous V$_S$ and V$_{2S}$ defects are unambiguously observed. The distributions of V$_S$ (white spheres) and V$_{2S}$ (yellow spheres) defects were shown in corresponding atomic mode (Fig. 2a). Enlarged STEM-ADF images clearly show that two edge-shared top and bottom S atoms in 1H-MoS$_2$ triangular prism escape simultaneously in V$_{2S}$, while only one S atom run away in V$_S$ (Fig. 2b), consistent with corresponding simulated images.

Figure 2c and d show the temperature dependence of the PL spectra upon excitation by light at 2.41 eV. The PL emissions of the A exciton (X$^A$) are blueshifted due to the crystal lattice contraction as the temperature decreases.[48] Above 150 K, PL emissions of A excitons are strikingly observed, and a new PL peak (X$^D$) appears under further cooling, which overlapps with the A exciton PL peak with similar PL intenisty, thus resulting in broaden line widths in the region of 150-75 K (Fig. 2d). However, the PL intensity of the A exciton drastically decreases below 75 K; the strong emission $X^D$ at ~1.74 eV is unambiguously observed, which has a Stokes shift of ~0.14 eV and an over twofold enhancement in the PL intensity as compared to the A exciton at 10 K. Figure 2e shows the absorption spectra of monolayer $MoS_2$ at 10 K. The A and B exciton absorption features are distinct; however, a new absorption feature at approximately 1.74 eV is strikingly detected. The splitting between the $X^A$ and $X^D$ is approximately 0.14 eV, similar to the PL emission spectra. The optically excited electron-hole pairs can be bound to excessive electrons by coulomb interactions in the heavily doped sample, giving rise to charged excitons (trions or Fermi polarons) in different doping regimes, leading to a PL emission below $X^A$, however, the Stokes shifts of trion and polarons should be ~40 meV,[4, 49-51] which is much smaller than that of $X^A$ and $X^D$ in our $MoS_2$. Therefore, the possibility of trions is ruled out. The strong emissions $X^D$ are attributed to the defect-bound excitons induced by chalcogen vacancies,[20, 37, 52] rather than charged excitons. The first-principles calculation results in Fig. 1d show that the defect states arising from S vacancy in monolayer $MoS_2$ appear at ~0.55 eV below the conduction band minimum (CBM), which is in agreement with the reported predicted results (~0.6 eV).[53-54] The band gap between defect states and CBM consist of the binding energy of A exciton and Stokes energy shift between the $X^A$ and $X^D$. Thus, the binding energy of $X^A$ is estimated to be approximately 0.41 eV, which is also consistent with the reported experimental results,[55-56] indicating that our calculated results of S vacancy defects are consistent with the experimental STEM-ADF and PL results.

The defect emission is still striking under nearly on-resonance excitation (1.96 eV) with the A exciton, whereas the Raman features of $MoS_2$ and Si overlap with the PL peak of the A exciton, particularly below 100 K, hampering to distinguish the A exciton feature (see Supplementary Information Fig. S3). The helicity parameter (*P*) of A exciton is approximately 80%, where $P = (I_{\sigma-} - I_{\sigma+})/(I_{\sigma-} + I_{\sigma+})$, and $I_{\sigma-}$ and $I_{\sigma+}$ are the intensities of the right-handed ($\sigma^-$) and left-handed ($\sigma^+$) circularly polarized PL. However, the defect emission is unpolarized upon excitation at 1.96 and 2.41 eV because off-resonance excitation results in the simultaneous occurrence of optical transitions in both the +K and –K valleys.[2]

The interaction of the magnetic moment with the magnetic field lifts the spin states, leading to splitting of the energy level. The defect states exhibit an unambiguous momentum dependence in the vicinity of the +K and –K valleys, implying the feasibility of controlling the defect pseudospin through the Zeeman effect under an external magnetic field. The exfoliated monolayer $MoS_2$ are used as the control sample to study the valley Zeeman splitting of A excitons. Figure 3a and 3b show the normalized polarization-resolved PL spectra of the defect-localized excitons and A excitons at selected magnetic fields. At zero magnetic field (middle), the right-handed (+K valley, red curve, $\sigma^-$) and left-handed (–K valley, blue curve, $\sigma^+$) PL emission completely overlap. However, the $\sigma^-$ ($\sigma^+$) PL peak shifts to a higher energy than that of $\sigma^+$ ($\sigma^-$) at -7 T (+7 T), indicating that valley Zeeman splitting of defect states takes place. Figure 3c shows the PL intensity difference ($\Delta PL$) in the normalized polarization-resolved PL spectra of defect states at ±7 T and 0 T, $\Delta PL = I(\sigma^-) - I(\sigma^+)$, where $I(\sigma^-)$ and $I(\sigma^+)$ are the $\sigma^-$ and $\sigma^+$ PL intensity, respectively. The $\Delta PL$ results further clearly reveals valley Zeeman splitting. At 0 T (black curve), $\Delta PL$ is almost equal to zero, however, $\Delta PL$ exhibits opposite values symmetric about 0 T when the magnetic field is ±7 T, with a positive (negative) value for +7 T (-7 T) below ~1.71 eV and an inverse trend above this value. This observation indicates that the spin DOF of defect states can be controlled by a magnetic field through the valley Zeeman effect. The "center of mass peak analysis" is used to obtain the PL peak position and valley splitting as a function of magnetic field.[57] The splitting as a function of $B$ is shown in Fig. 3d, yielding a *g*-factor of -6.2 for defect excitons which is enhanced by ~50% as compared to that of A exciton in exfoliated monolayer $MoS_2$ (*g* = -4.2).

Further insights into the origin of valley Zeeman splitting can be obtained from the calculated electronic band structures of monolayer $MoS_2$ with and without defects. The valence band maximum (VBM) of monolayer $MoS_2$ without S vacancies is only comprised of the $d_{x^2-y^2}$ and $d_{xy}$ orbitals of Mo with $m = ±2$ in the ±K valley, while the CBM consists of the $d_{z^2}$ orbital of Mo with $m = 0$ (Fig. 4a); the valley magnetic moments are $±(m_0/m^*_{v/c})\mu_B$ for the –K and +K points, therefore, the *g*-factor is approximately -4.[52, 58] However, five *d*-orbitals of the Mo atom contribute to the defect states and VBM in monolayer $MoS_2$ with $V_S$ and $V_{2S}$ vacancies (Fig. 4b and Supplementary Information Fig. S4), including the $d_{zx}$ and $d_{zy}$ orbitals ($m = ±1$), the $d_{x^2-y^2}$ and $d_{xy}$ orbitals ($m = \pm 2$) and the $d_{z^2}$ orbital ($m = 0$), whereas the CBM is still composed of the $d_{z^2}$ orbital. Thus, the orbital magnetic moment leads to no shift in the conduction band, but a shift in the valence band edge and defect states at the –K and +K points.

The splitting of defect-bound excitons can be understood as a result of the interaction of the spin, orbital and valley magnetic moments with the magnetic field, as illustrated in Fig. 5. Two degenerate but inequivalent valleys (±K valley) are associated with each other via the time-reversal symmetry in monolayer MoS$_2$, giving rise to opposite valley pseudospins in the ±K valleys.[34] Upon excitation by 2.41 eV $\sigma^-$ circularly polarized light, off-resonance excitation simultaneously occurs in both the +K and –K valleys (Fig. 5a), as a consequence no valley-spin polarization is observed.[2] Defects trap carriers in real space and provide new recombination channels. The valley lifetime ($\gamma_1$) in 2D TMDs is at least tens of nanoseconds at least,[59] however, the charge transfer ($\gamma_2$) from excited states to defect states occurs within ~1 ps, which is at least three orders of magnitude shorter than the valley spin lifetime.[60] Thus, excited electrons quickly relax to defect states between two bands with the same spin states (Fig. 5b). Under an out-of-plane positive magnetic field, the spin-degenerate bands are lifted. The contributions from the spin magnetic moment and orbital magnetic moment of the $d_{zx}$ and $d_{zy}$ orbitals are the same for the defect and valence band shifts (Fig. 5c); thus these orbitals do not contribute to the valley Zeeman splitting.[52, 57] However, the components of the $d_{x^2-y^2}$ and $d_{xy}$ orbitals in the defect and valence band edges are significantly different (Fig. 4b and 5c); thus these orbitals lead to different band shifts of the defect and valence states with the same shift direction in the same valley, resulting in an optical resonance shift of $-5\mu_B B/3$ in the $-K$ valley. The $V_S$ defect ($m_d^*$) and valence ($m_v^*$) band masses are corrected in the $-K$ and $+K$ valleys, estimated to be $3.89m_0$ and $0.53m_0$ through first-principles calculations, resulting in a valley magnetic moment of $\pm(m_0/m_{v/d}^*)\mu_B$ for the valence and defect states, where $\mu_B$ is the Bohr magneton.[52] The effective mass of the electron in the defect band edge is over 7 times that of the hole in the valence band edge, resulting in a larger shift of the defect band compared to the valence band, which leads to a significant optical resonance shift in the +K and –K valleys, $\Delta E_{v(-K)} = [(m_0/m_d^*) - (m_0/m_v^*)]\mu_B B$ (Fig. 5c). Thus, the valley Zeeman splitting ($\Delta E$) of defect states is mainly determined by the atomic orbital and valley magnetic moments, which is given by $\Delta E = -2\{5/3 - [(m_0/m_d^*) - (m_0/m_v^*)]\}\mu_B B$; thus, the calculated effective $g$-factor for $V_S$ defect states is approximately -6.6, and $V_{2S}$ defect states have a $g$-factor of approximately -6.3 (See Supplementary Information Fig. S4), both of which are consistent with the experimental results ($g$ = -6.2). Therefore, the enhanced $g$-factor for defect-bound excitons originates from the increased effective electron mass of defect states and $d$-orbital magnetic moment, indicating that the $g$-factor is independent of defect density

In summary, we have demonstrated valley pseudospin of defect states in CVD-grown monolayer $MoS_2$. Five $d$-orbitals of the Mo atom and the large effective mass of electrons in defect states lead to enhanced valley Zeeman splitting and effective $g$-factor. Our results are not limited to monolayer $MoS_2$ and can be observed in a variety of TMDs, including $WSe_2$, $MoSe_2$ and $WS_2$. These features highlight the potential to control the spin through defects toward novel quantum information devices and quantum spintronic and valleytronic devices.[30-33]

**ASSOCIATED CONTENT**

**Supporting Information**

The Supporting Information is available free of charge on the ACS Publications website

S1: Band structures of monolayer $MoS_2$ with one S divacancy ($V_{2S}$) defect and without defect.

S2: Raman feature of monolayer $MoS_2$.

S3: On-resonance excitation with the A exciton

S4: Five $d$-orbital components in the electronic band structures of monolayer $MoS_2$ with one $V_{2S}$ defect.

**Methods**

**Sample Preparation:** Monolayer $MoS_2$ was grown by CVD using S powder (99.999%, Ourchem) and $MoO_3$ (99.99%, Ourchem) as sources, PTAS as a seeding promoter, pieces of a 300 nm $SiO_2$/Si wafer as substrates, and high-purity inert argon as the carrier gas. The temperatures for the S and $MoO_3$ sources were elevated from room temperature to 160 and 650 °C, respectively.

**Optical Spectroscopy Measurement:** The PL, Raman, and reflection signals were recorded by a Witec Alpha 300R Plus confocal Raman microscope, coupled with a 7 T superconducting magnetic field and a closed cycle optical cryostat (10 K) on an XY scanning stage. A long working distance 50× objective (Olympus NA = 0.55) was used for the low-temperature PL and reflection measurements. Polarization-resolved PL spectra were obtained under circular polarization excitation of 4 mW at both 1.96 eV and 2.41eV, collected by the same objective, and passed through a 1/4λ waveplate and an analyzer into a spectrometer and a CCD camera.

**STEM Characterization and Image Simulation**: Monolayer $MoS_2$ flakes were transferred onto a TEM grid from the $SiO_2$/Si substrates through a PMMA films. An aberration-corrected JEOL ARM-200F with a cold field emission gun and an ASCOR probe corrector at 60 kV was used for STEM-ADF imaging. The convergence semiangle of the probe was ≈30 *mrad*. STEM-ADF images were collected using a half-angle range from ≈85 to 280 *mrad*. The QSTEM package was used for image simulations by assuming an aberration-free probe and an ≈1 Å source size to give a probe size of ≈1.2 Å.

**First-Principles Calculations:** Fully relativistic calculations within density functional theory are employed using the Quantum-ESPRESSO package. We choose ultrasoft pseudopotentials and the generalized gradient approximation (Perdew-Burke-Ernzerhof parametrization) of the exchange correlation functional in present calculations. In the calculation, a 4 × 4 supercell is used. Due to Brillion zone folding, K and -K points of the 1 × 1 unit cell are folded onto the K and -K points of the 4 × 4 supercell, respectively. A 15 Å thick vacuum layer is adopted to avoid artificial interaction because of the periodic boundary conditions. A high cutoff energy of 544 eV and a precise 4 × 4 ×1 for the k-point sampling are used. The structural optimization is continued until the residual forces have converged to less than 2.6 × $10^{-3}$ eV/ Å and the total energy to less than $1.4 \times 10^{-4}$ eV. The spin-orbit interaction is considered for the band structure calculation. To model one $V_S$ or double $V_{2S}$ defect in monolayer $MoS_2$, one S atom or two S atoms in the same column are removed in $4 \times 4$ supercell.

**AUTHOR INFORMATION**

**Corresponding Author**

\* Bo Peng: bo_peng@uestc.edu.cn

\* Xiaoxu Zhao: xiaoxu_zhao@u.nus.edu

\* Yingchun Cheng: iamyccheng@njtech.edu.cn

**ORCID**

Bo Peng: 0000-0001-9411-716X

**Author contributions**

B.P. developed the concept, designed the experiment. B.P., Y.Q.W., L.J.D, Z.L., Y.L. prepared the manuscript. Y.Q.W., X.L. performed the PL and Raman measurements. Y.W., Y.Y. synthesized the monolayer $MoS_2$. Q.L.W., W.H., Y.C.C. contributed to first-principles calculations. X.X.Z, K.P.L, S.J.P contributed to the STEM measurements. L.B., L.Z., H.P.L., H.Y.C, P.H.Z, L.B.Z. discussed the mechanism of valley Zeeman splitting.

**Notes**

The authors declare no competing financial interest.

**ACKNOWLEDGMENTS**

We acknowledge financial support from National Natural Science Foundation of China (51602040, 51872039, 51525202 and 51902098), Science and Technology

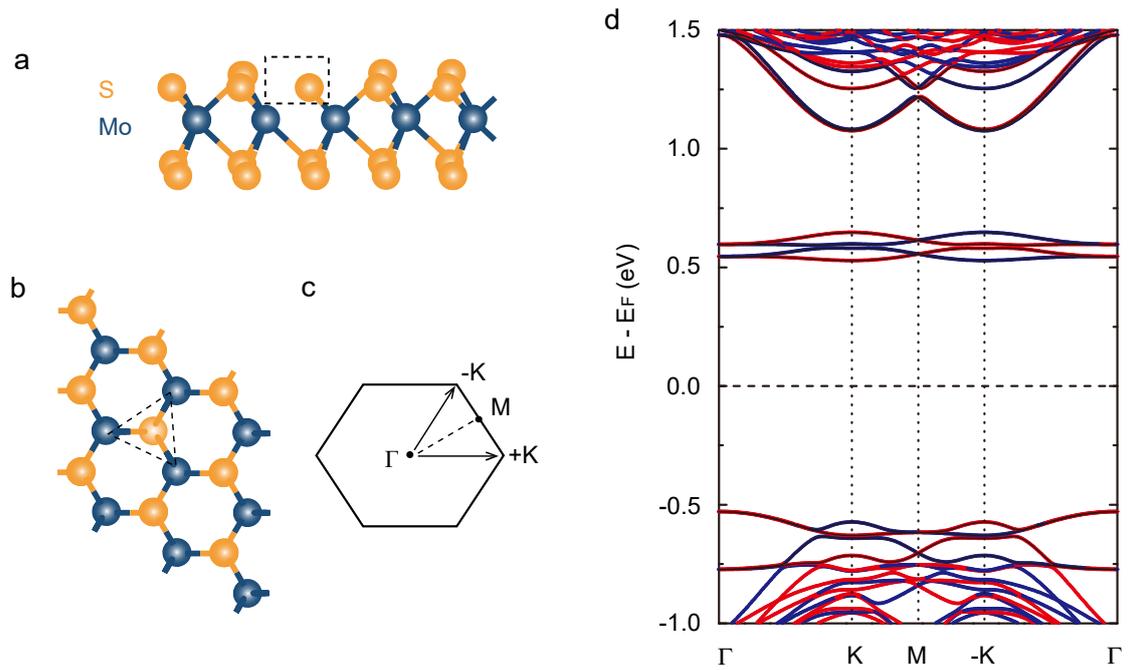

**Fig. 1. Band structures of monolayer MoS$_2$ with one V$_S$ defect. (a, b)** Side and top views of the monolayer 1H-MoS$_2$ structure with one V$_S$ defect. **(c)** Brillouin zone of monolayer MoS$_2$. **(d)** Electronic band structure of monolayer MoS$_2$ with one V$_S$ defect. Blue (red) curves represent spin-up (spin-down) states in the +K (-K) valley.

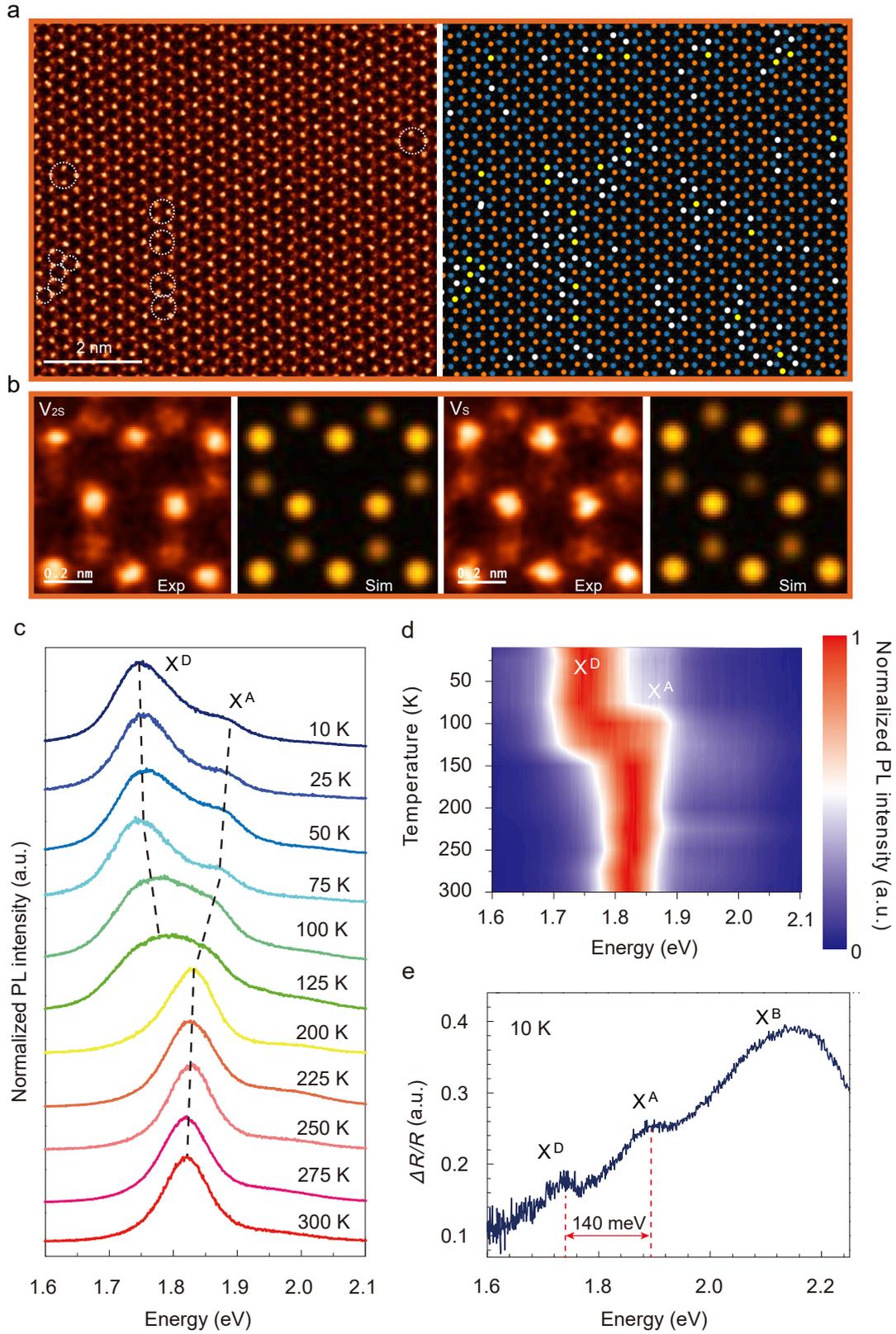

**Fig. 2. Strong defect emission from $V_S$ and $V_{2S}$ defects in monoalyer $MoS_2$. (a)** Atomic-resolution STEM-ADF images of 1H-$MoS_2$ monolayer with $V_S$ and $V_{2S}$

(white dashed circle) defects. Corresponding atomic mode showing the distribution of $V_S$ and $V_{2S}$ defect, which are depicted as white and yellow spheres. **(b)** Enlarged STEM-ADF image and corresponding simulated image revealing $V_S$ and $V_{2S}$ defect. The scale bare in b is 0.2 nm. **(c, d)** PL spectra of monolayer $MoS_2$ and corresponding two dimensional image as a function of temperature upon excitation by 2.41 eV light. A strong defect emission ($X^D$) accompanied with A exciton emission ($X^A$) is observed below 100 K. **(e)** Absorption spectra of the corresponding monolayer $MoS_2$ at 10 K, indicating an unambiguous defect absorption features, in addition to the A exciton and B exciton ($X^B$) features.

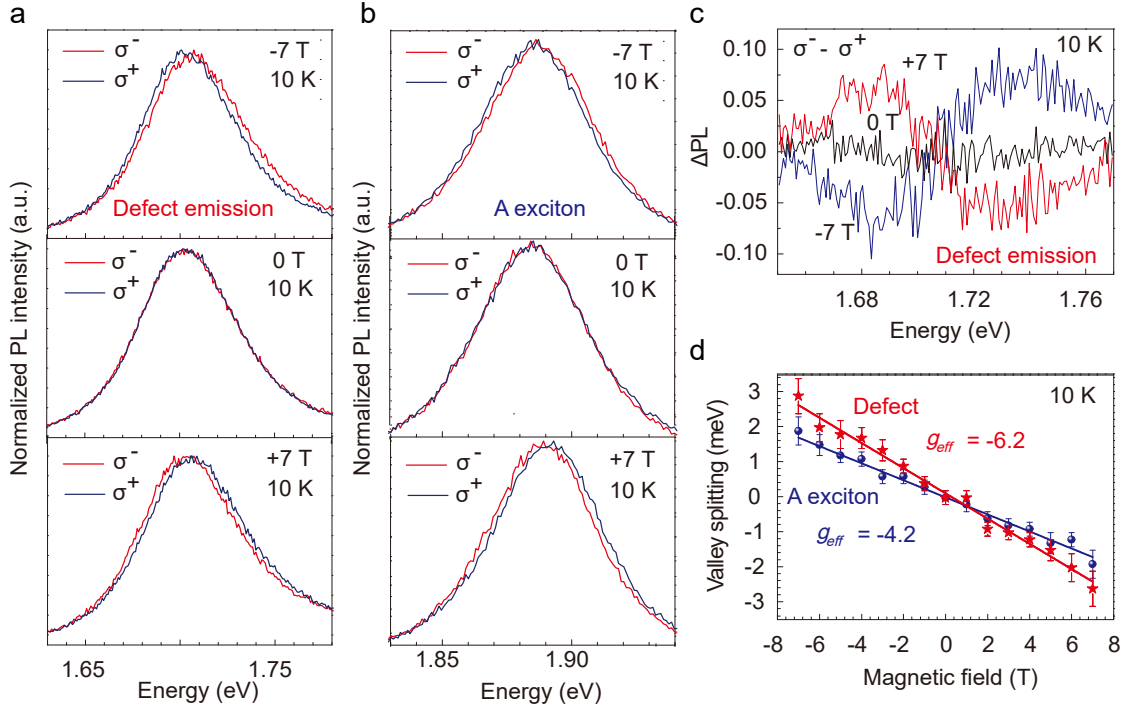

**Fig. 3. Valley Zeeman splitting of defect emission. (a, b)** Normalized raw polarization-resolved PL spectra of defect-localized excitons and A exciton (exfoliated MoS$_2$) in magnetic field of 0 and ±7 T at 10 K. The σ⁻ and σ⁺ PL spectra completely overlap at 0 T, but they split at +7 T and −7 T with opposite shift direction. **(c)** Difference in the normalized σ⁻ and σ⁺ PL components at 0 and ±7 T, which show no difference at 0 T but striking opposite trends at ±7 T, strongly manifesting that valley Zeeman splitting of defect states occurs. **(d)** Valley splitting as a function of the magnetic field at 10 K, giving an effective *g*-factor of -6.2 (defect-localized excitons) and -4.2 (A excitons). The PL peak positon and valley splittings as a function of magnetic field were obtained through "center of mass peak analysis" (Ref. 48).

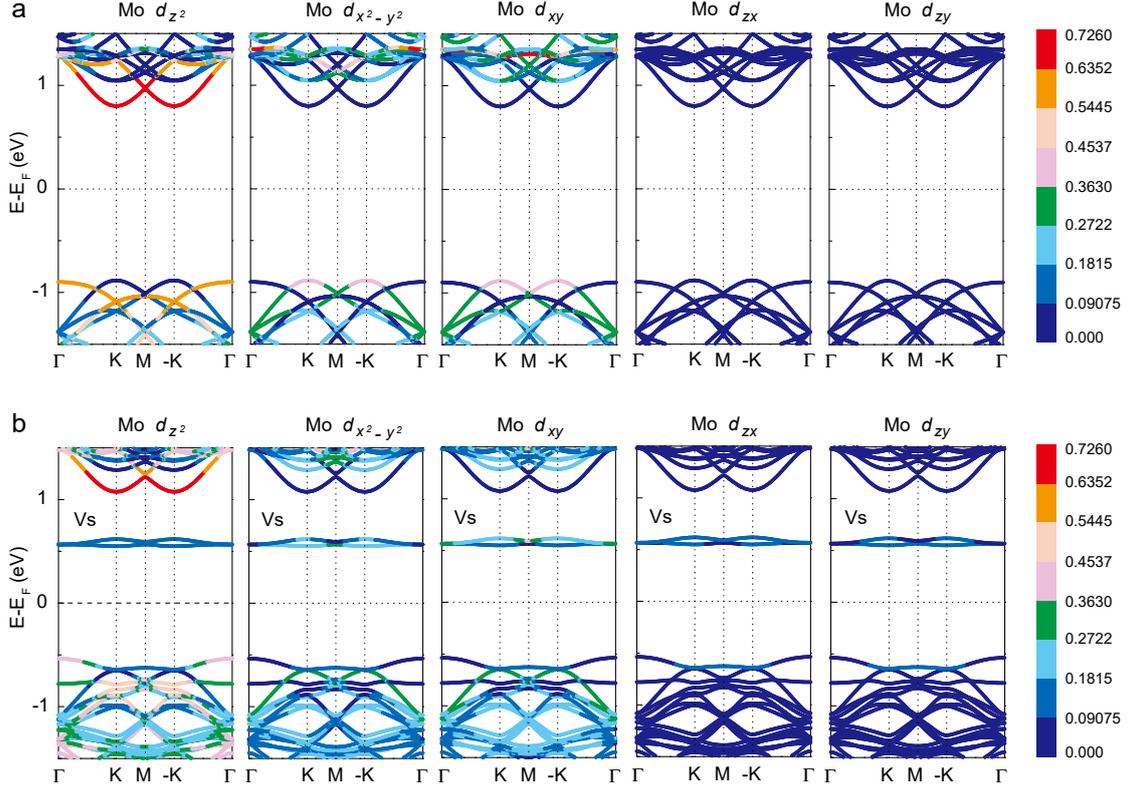

**Fig. 4. Origin of valley Zeeman splitting of defect states. (a, b)** Five *d*-orbital components in the electronic band structures of monolayer MoS$_2$ without defects **(a)** and with one V$_S$ defect **(b)**. The conduction band edges of monolayer MoS$_2$ with and without defects consist of the $d_{z^2}$ (*m*=0) orbitals. The valence band edges of monolayer MoS$_2$ without defect in the vicinity of the ±K valleys are mainly composed of the $d_{x^2-y^2}$ (*m*=±2) and $d_{xy}$ (*m*=±2) orbitals. However, the valence and defect band edges monolayer MoS$_2$ with one single S vacancy are composed of all five *d*-orbitals.

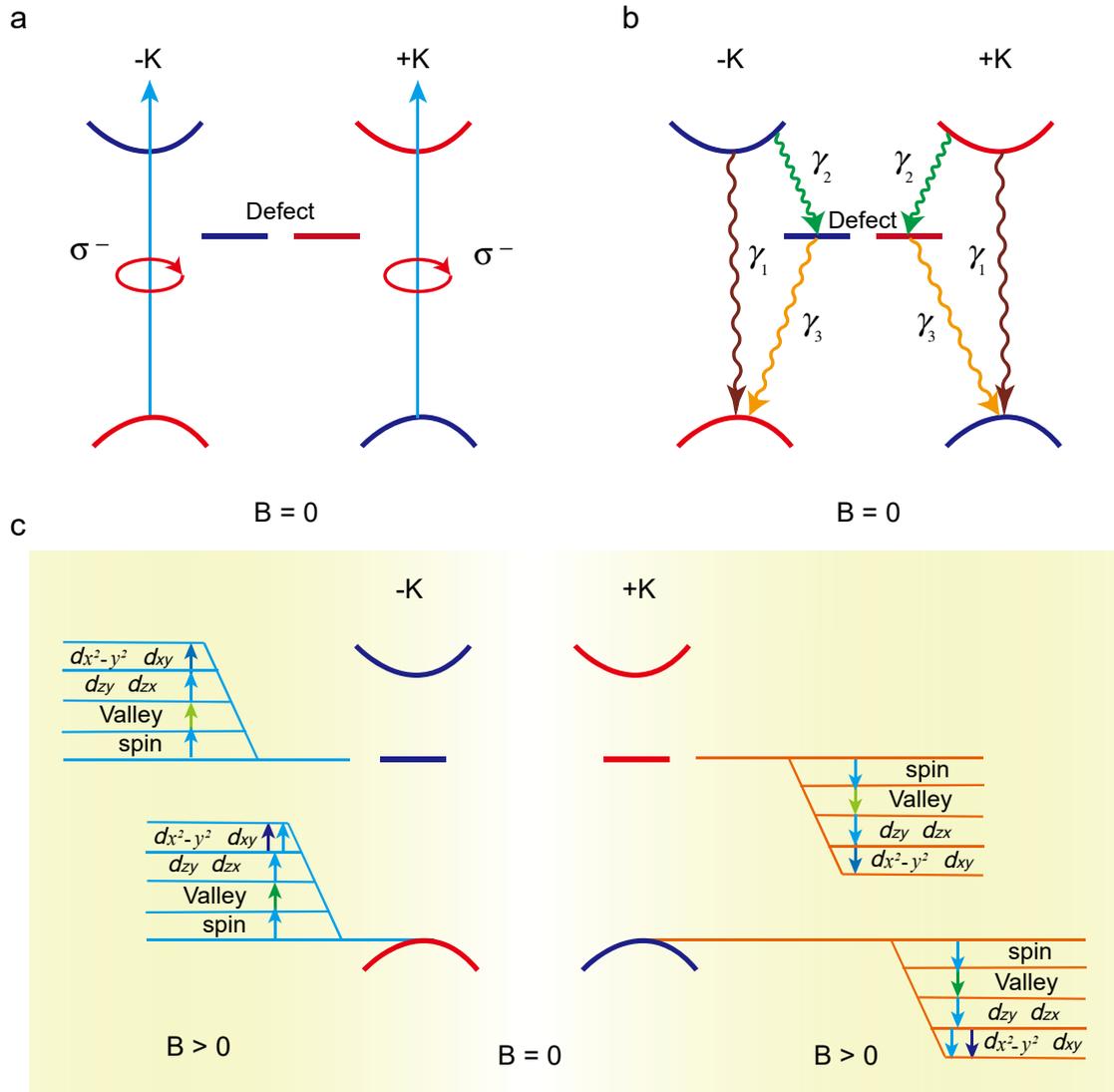

**Fig. 5. Valley Zeeman splitting of defect-bound excitons under a magnetic field.**

**(a)** Off-resonance excitation by right-hand polarized light at 2.41 eV simultaneously occur in the –K and +K valleys. **(b)** Corresponding charge carrier relaxation process upon off-resonance excitation. **(c)** Valley Zeeman splitting under a magnetic field as a result of contributions from the spin magnetic moment, valley magnetic moment and $d$-orbital magnetic moment of the band edge of defect and valence states.